\documentstyle[12pt]{article}
\title{\bf Casimir effect for a spherical shell\\ in de Sitter
spacetime  with  signature change}
\author{M. R. Setare \thanks{e-mail:rezakord@ipm.ir}\hspace{5mm}; F. Darabi \thanks{e-mail:f.darabi@azaruniv.edu}
\\
{\small Physics Dept. Inst. for Studies in Theo. Physics and
Mathematics (IPM), 19395-5531, Tehran, Iran .}\\
{\small Department
of Physics, Azarbaijan University of Tarbiat Moallem, 53714-161
Tabriz, Iran .} }
\begin{document}
\maketitle \vspace{15mm}

\begin{abstract}
The Casimir stress on a spherical shell in de Sitter signature
changing background for massless scalar field satisfying Dirichlet
boundary conditions on the shell is calculated. The Casimir stress
is calculated for inside and outside of the shell with different
backgrounds corresponding to different metric signatures and
cosmological constants. An important contribution appears due to
signature change which leads to a transient rapid expansion of the
bubbles in this background.
\end{abstract}

\section{Introduction}

The Casimir effect is regarded as one of the most striking
manifestation of vacuum fluctuations in quantum field theory. The
presence of reflecting boundaries alters the zero-point modes of a
quantized field, and results in the shifts in the vacuum
expectation values of quantities quadratic in the field, such as
the energy density and stresses. Therefore, the Casimir effect can
be viewed as the polarization of the vacuum by boundary conditions
or geometry. In particular, vacuum forces arise acting on the
constraining boundaries. The particular features of these forces
depend on the nature of the quantum field, the type of spacetime
manifold and its dimensionality, the boundary geometries and the
specific boundary conditions imposed on the field. Since the
original work by Casimir in 1948 \cite{Casi48} many theoretical
and experimental works have been done on this problem
\cite{db,Most97,Plun86,Lamo99,Bord01,Kirs01,Schw,setare}.

The time dependence of boundary conditions or geometries, the
so-called dynamical Casimir effect, is also a new element which
has to be taken into account. In particular, in \cite{set} the
Casimir effect has been calculated for a massless scalar field
satisfying Dirichlet boundary conditions on the spherical shell in
de Sitter space. The Casimir stress is calculated for inside and
outside of the shell with different backgrounds corresponding to
different cosmological constants.

On the other hand, signature changing spacetimes have recently
been of particular importance as specific geometries with
interesting physical effects. The initial idea of signature change
is due to Hartle, Hawking and Sakharov \cite{HHS} which makes it
possible to have a spacetime with Euclidean and Lorentzian regions
in quantum gravity. It has been shown that the signature change
may happen even in classical general relativity \cite{CSC}. The
issue of propagation of quantum fields on signature-changing
spacetimes has also been of some interest \cite{D}. For example,
Dray {\it et al} have shown that the phenomenon of particle
production can happen for propagation of scalar particles in
spacetime with heterotic signature. They have also obtained a rule
for propagation of massless scalar fields on a two dimensional
spacetime with signature change.

The Casimir effect has also recently been studied in a signature
changing spacetime and shown that there is a non-vanishing
pressure on the hypersurface of signature change \cite{FM}.
Motivated by this new element in studying the Casimir effect we
have paid attention to study such a non-trivial effect in a model
of a spherical shell in de Sitter space with different
cosmological constants and metric signatures inside and outside.
Our aim is to understand the possible role of such a non-trivial
effect in the time evolution of the bubbles at early universe with
false/true vacuum and Euclidean/Lorentzian metrics,
outside/inside.

\section{Scalar Casimir effect for a sphere in de Sitter space}

  The Casimir force due to fluctuations of a free massless
  scalar field satisfying Dirichlet boundary conditions on a spherical
  shell in Minkowski space-time has been studied in \cite{mil}. The two-point Green's
  function $G(x,t;x',t')$ is defined as the vacuum expectation
  value of the time-ordered product of two fields
  \begin{equation}
  G(x,t;x',t')\equiv-\imath<0|T\Phi(x,t)\Phi(x',t')|0>.
  \label{1}
  \end{equation}
  It has to satisfy the Dirichlet boundary conditions on the
  shell:
  \begin{equation}
  G(x,t;x',t')|_{|x|=a}=0,
  \label{2}
  \end{equation}
  where $a$ is radius of the spherical shell. The stress-energy tensor
  $T^{\mu\nu}(x,t)$ is given by
  \begin{equation}
  T^{\mu\nu}(x,t)\equiv\partial^{\mu} \Phi(x,t)\partial^{\nu} \Phi(x,t)-
  \frac{1}{2}\eta^{\mu\nu} \partial_{\lambda} \Phi(x,t)\partial^{\lambda}
  \Phi(x,t).
  \label{3}
  \end{equation}
  The radial Casimir force per unit area $\frac{F}{A}$ on the
  sphere, called Casimir stress, is obtained from the radial-radial component of the vacuum
  expectation value of the stress-energy tensor:
  \begin{equation}
  \frac{F}{A}=\langle0|T^{rr}_{in}-T^{rr}_{out}|0\rangle|_{r=a}.
  \label{4}
  \end{equation}
  Taking into account the relation (1) between the vacuum expectation value
  of the stress-energy tensor $T^{\mu\nu}(x,t)$ and the Green's
  function at equal times $ G(x,t;x',t)$ we obtain
  \begin{equation}
  \frac{F}{A}=\frac{i}{2}[\frac{\partial}{\partial r}\frac{\partial}{\partial
  r'}G(x,t;x',t)_{in}-\frac{\partial}{\partial r}\frac{\partial}{\partial
  r'}G(x,t;x',t)_{out}]|_{x=x',|x|=a}.
  \label{5}
  \end{equation}
  One may use of the above flat space calculation in de Sitter space-time by taking the de Sitter metric in conformally flat
form
  \begin{equation}
  ds^{2}=\Omega(\eta)[d\eta^{2}-\sum_{\imath=1}^{3}(dx^{\imath})^{2}],
  \label{6}
  \end{equation}
  where $\Omega(\eta)=\frac{\alpha}{\eta}$ and $\eta$ is the conformal time
  \begin{equation}
  -\infty <  \eta < 0.
  \label{7}
  \end{equation}
  Assuming a canonical quantization of the scalar field, the conformally transformed quantized
   scalar field in de Sitter space
 is given by
\begin{equation}
  \bar\Phi(x,\eta)=\sum_{k}[a_{k}
  \bar u_{k}(\eta,x)+a_{k}^{\dagger}\bar u_{k}^{\ast}(\eta,x)],
  \label{8}
  \end{equation}
  where $a_{k}^{\dagger}$ and
  $a_{k}$ are  creation and annihilation operators respectively
  and  the vacuum states associated with the modes $\bar u_{k}$
defined by
  $a_{k}|\bar 0\rangle=0 $, are called conformal vacuum.
  Given the flat space Green's function(1), we obtain
  \begin{equation}
 \bar G=-i\langle\bar{0}|T \bar \Phi(x,\eta)\bar
 \Phi(x',\eta^{'})|\bar{0}\rangle=\Omega^{-1}(\eta)\Omega^{-1}(\eta^{'})G,
 \label{9}
 \end{equation}
where $\bar\Phi(x,\eta)=\Omega^{-1}(\eta)\Phi(x,\eta)$ has been
used. Therefore, using Eqs.(\ref{4}), (\ref{5}) and (\ref{9}) we
obtain the total stress on the sphere in de Sitter space as
\begin{equation}
(\frac{\bar F}{A})=\frac{\eta^{2}}{\alpha^{2}} \frac{F}{A}.
\label{10}
\end{equation}

\section{Spherical shell with different vacua and signatures inside and outside}

We assume different vacua inside and outside, corresponding to
different $\alpha_{in}$ and $\alpha_{out}$ for the Lorentzian
metric (\ref{6}) and use the following relation for the stress on
the shell due to boundary conditions in flat spacetime \cite{Schw}
\begin{equation}
\frac{F}{A}=\frac{-1}{4\pi a^{2}}\frac{\partial E}{\partial a},
\label{11}
\end{equation}
where the Casimir energy $E$ is the sum of Casimir energies
$E_{in}$ and $E_{out}$ for inside and outside of the shell. The
corresponding relation in de Sitter space is given by \cite{set}
\begin{equation}
\frac{\bar{F}}{A} = \frac{-1}{4\pi
a^{2}}\frac{\partial\bar{E}}{\partial a}=\frac{\eta^{2}}{8\pi
a^{4}}(\frac{c_{1}}{\alpha_{in}^{2}}+\frac{c_{2}}{\alpha_{out}^{2}}).
\label{12}
\end{equation}
where we have used the renormalized total zero-point energy
\begin{equation}
\bar{E}=\frac{\eta^{2}}{2a}(\frac{c_{1}}{\alpha_{in}^{2}}+\frac{c_{2}}{\alpha_{out}^{2}}).
\label{13}
\end{equation}
in which $c_{1} = 0.008873$, $c_{2} = -0.003234$.

Now, we obtain the pure effect of vacuum polarization due to the
gravitational field without any boundary conditions in Euclidean
(outside) region with the following metric
\begin{equation}
ds^{2}=-\Omega(\eta)[d\eta^{2}+\sum_{\imath=1}^{3}(dx^{\imath})^{2}].
\label{14}
\end{equation}
To this end, we calculate the renormalized stress tensor for the
massless scalar field in de Sitter space with Euclidean signature.
One may use \cite{db}
\begin{equation}
\langle0|T^{\nu}_{\mu}[g_{kl}]|0\rangle|_{ren}=(\tilde{g}/g)^{1/2}\langle0|T^{\nu}_{\mu}[\tilde{g}_{kl}]|0\rangle|_{ren}-
\frac{1}{2880\Pi^2}[{\frac{1}{6}}\vspace{2mm} ^{(1)}H^{\nu}_{\mu}-
^{(3)}H^{\nu}_{\mu}], \label{15}
\end{equation}
where $\tilde{g}_{kl}$ is the flat Euclidean metric for which
$\langle0|T^{\nu}_{\mu}[\tilde{g}_{kl}]|0\rangle|_{ren}=0$, and
$$
^{(1)}H^{\nu}_{\mu}=0,
$$
$$
^{(3)}H^{\nu}_{\mu}=\frac{3}{\alpha^4}\delta^{\nu}_{\mu}.
$$
We then obtain
\begin{equation}
\langle0|T^{\nu}_{\mu}[g_{kl}]|0\rangle|_{ren}=\frac{1}{960 \Pi^2
\alpha^4}\delta^{\nu}_{\mu},
\label{16}
\end{equation}
which is exactly the same result for the Lorentzian case
\cite{db}. Therefore, the corresponding effective radial pressures
for the Euclidean (outside) and Lorentzian (inside) regions with
$\alpha_{out}$ and $\alpha_{in}$, due to pure effect of
gravitational vacuum polarization without any boundary condition,
are given respectively by
$$
P_{out}^{E}=-\langle0|T^{r}_{r}[g_{kl}]|0\rangle|_{ren}=-\frac{1}{960
\Pi^2 \alpha^4_{out}}
$$
$$
P_{in}^{L}=-\langle0|T^{r}_{r}[g_{kl}]|0\rangle|_{ren}=-\frac{1}{960
\Pi^2 \alpha^4_{in}}.
$$
The corresponding gravitational pressure on the spherical shell is
then given by
\begin{equation}
P_G = P_{in}^{L}-P_{out}^{E} =
-\frac{1}{960\pi^{2}}(\frac{1}{\alpha_{in}^{4}}-\frac{1}{\alpha_{out}^{4}}).
\label{17}
\end{equation}
We now proceed to calculate the stress due to the boundary effects
$P_B$. To this end, we make maximum use of the results obtained in
\cite{set}. The stress on the shell due to boundary effects for
the Lorentzian metric (\ref{6}) has been obtained as (\ref{12}).
In signature changing case we have correspondingly
\begin{equation}
\left(\frac{\tilde{F}}{A}\right)_{L-E}=\langle0|T_{rr}|0\rangle|^{L}_{in}-\langle0|T_{rr}|0\rangle|^{E}_{out},
\label{18}
\end{equation}
where
\begin{equation}
\langle0|T_{\mu \nu}|0\rangle|=\sum_{\alpha}T_{\mu
\nu}\{\Phi_{\alpha}, \Phi_{\alpha}^{*}\}. \label{19}
\end{equation}
The detailed calculations show that the only difference between
the Lorentzian case and signature changing one appears as follows
\begin{equation}
\left(\frac{\tilde{F}}{A}\right)_{L-E}=\left(\frac{\tilde{F}}{A}\right)_{L}+\partial_{\eta}\partial^{\eta}
\{\Phi_{\alpha_{in}},
\Phi_{\alpha_{in}}^{*}\}_{L}-\partial_{\eta}\partial^{\eta}
\{\Phi_{\alpha_{out}}, \Phi_{\alpha_{out}}^{*}\}_{E}.
\label{20}
\end{equation}
The scalar field $\Phi(r, \theta, \eta)$ in the Lorentzian de
Sitter space satisfies
\begin{equation}
(\Box+\xi R)\Phi(r, \theta, \eta)=0,
\label{21}
\end{equation}
where $\Box$ is the Laplace-Beltrami operator for the de Sitter
metric, and $\xi$ is the coupling constant. For conformally
coupled field in four dimension $\xi=\frac{1}{6}$, and R , the
Ricci scalar curvature, is given by
\begin{equation}
R=12\alpha^{-2}.
\end{equation}
Taking into account the separation of variables as
\begin{equation}
\Phi_L(r,\theta,\eta)=A(r)B(\theta)T_L(\eta),
\label{22}
\end{equation}
for the inside Lorentzian domain with
\begin{equation}
T_L(\eta)=\exp^{-i\omega\eta}, \label{23}
\end{equation}
the corresponding Euclidean $\eta$-dependence takes on the form
\begin{equation}
T_E(\eta)=\exp^{-\omega\eta},
\label{24}
\end{equation}
for the scalar field to be normalizable in $\eta$. Inserting
Eqs.(\ref{23}) and (\ref{24}) into Eq.(\ref{20}) for
$\Phi_L(r,\theta,\eta)$ and $\Phi_E(r,\theta,\eta)$, respectively
we obtain
\begin{equation}
\left(\frac{\tilde{F}}{A}\right)_{L-E}=\frac{\eta^{2}}{8\pi
a^{4}}(\frac{c_{1}}{\alpha_{in}^{2}}+\frac{c_{2}}{\alpha_{out}^{2}})
+\frac{\alpha_{out}^2}{\eta^2}(8\omega^2e^{-2\omega\eta}).
\label{25}
\end{equation}
Taking into account the gravitational pressure on the shell we
obtain the total result
\begin{equation}
P=P_G+P_B=-\frac{1}{960\pi^{2}}(\frac{1}{\alpha_{in}^{4}}-\frac{1}{\alpha_{out}^{4}})+
\frac{\eta^{2}}{8\pi
a^{4}}(\frac{c_{1}}{\alpha_{in}^{2}}+\frac{c_{2}}{\alpha_{out}^{2}})
+\frac{\alpha_{out}^2}{\eta^2}(8\omega^2e^{-2\omega\eta}).
\label{26}
\end{equation}
Inserting $\alpha^2=\frac{3}{\Lambda}$ we obtain the total
pressure in terms of the cosmological constants
\begin{equation}
P=-\frac{1}{2880\pi^{2}}(\Lambda_{in}^{2}-\Lambda_{out}^{2})
+\frac{\eta^{2}}{24\pi
a^{4}}(c_{1}\Lambda_{in}+c_{2}\Lambda_{out})+\frac{3}{\Lambda_{out}\eta^2}(8\omega^2
e^{-2\omega \eta} ).
\label{27}
\end{equation}
The first two terms are not new ( see \cite{set}), but the third
one shows a nontrivial effect due to signature changing
background. The first term $P_G$ which is pure gravitational
effect without boundary is $\eta$-independent, but is sensitive to
the initial values of $\Lambda_{in}$ and $\Lambda_{out}$. In the
inflationary interesting case, namely the true vacuum inside and
false vacuum outside, i.e. $\Lambda_{in}<\Lambda_{out}$, this
pressure is always repulsive. However, the last two terms are
$\eta$-dependent resulting from boundary effects. So, one has to
proceed cautiously. Let us assume
\begin{equation}
c_{1}\Lambda_{in}+c_{2}\Lambda_{out}>0.
\label{28}
\end{equation}
This leads the second term to be an ever increasing repulsive
pressure with time $\eta$, so that the pressure due to first two
terms is always positive. On the other hand, taking
\begin{equation}
c_{1}\Lambda_{in}+c_{2}\Lambda_{out}<0,
\label{29}
\end{equation}
with $|c_1| > |c_2|= - c_2$, the pressure due to first two terms
may be either negative or positive. If this pressure is initially
positive, the initial repulsion of the expanding bubble will be
stopped at an specific time $\eta$ and then a negative pressure
leads to a contracting shell ending up with a collapse of the
bubble. If on the other hand, this pressure is initially negative
it remains negative forever. Now, we consider the third term which
is of particular importance in the present work. This term acts as
dynamically decaying pressure which is very large at the beginning
$\eta \simeq 0$, but is vanishing at late times $\eta\gg 0$.
Moreover, it is positive provided the false vacuum
$\Lambda_{out}>0$, which is reasonable in the context of
inflationary models. The existence of this new dynamical pressure
may play an important role in the inflationary phase at early
universe.

In case of (\ref{29}), the general behavior of the bubble is not
affected by the third term because it is transient and its
presence merely alters the time $\eta$ of the contracting phase
and the final collapse of the bubble.

However, in case of (\ref{28}) which is of particular interest in
inflationary phase at early universe, all contributions are
positive and the bubble initially ($\eta \simeq 0$) experiences a
huge repulsion mainly due to the third term. At late times
$\eta\gg0$ the third term almost vanishes, and the second term
plays the important role in the late time repulsion of the bubble.

\section{Conclusion}

We have studied the Casimir effect for spherical bubbles with
different vacua and metric signatures inside and outside,
corresponding to de Sitter metrics. The metrics inside and outside
are taken Lorentzian and Euclidean, respectively. The case of
different vacua in a Lorentzian de Sitter spacetime has already
been studied in \cite{set}. In the present work we have shown how
those results are affected in a signature changing background. It
is shown that a transient term appears due to signature change,
which in comparison to the Lorentzian case ( true vacuum inside
and false vacuum outside ) results initially in an extra rapid
expansion of the bubble. This extra rapid expansion vanishes at
late times and the Lorentzian results are then recovered. It
reveals that the presence of signature change may cause to more
rapid transient expansion of bubbles than is predicted by
Lorentzian case in \cite{set}. This may have some important
impacts on the inflationary phase at early universe.

\section*{Acknowledgment}

This work has been financially supported by the Research
Department of Azarbaijan University of Tarbiat Moallem, Tabriz,
Iran.


\begin{thebibliography}{99}

\bibitem{Casi48} H. B. G. Casimir, Proc. K. Ned. Akad. Wet. {\bf
51}, 793 (1948).
\bibitem{db} N. D. Birrel and P. C. W. Davies, {\it Quantum Fields in
Curved Space} (Cambridge: Cambridge University Press, 1982).
\bibitem{Most97}  V. M. Mostepanenko and N. N. Trunov, {\it The
Casimir Effect and Its Applications} (Clarendon, Oxford, 1997).
\bibitem{Plun86}  G. Plunien, B. Muller, and W. Greiner, Phys. Rep.
{\bf 134}, 87 (1986).
\bibitem{Lamo99} S. K. Lamoreaux, Am. J. Phys. {\bf 67}, 850 (1999).
\bibitem{Bord01} M. Bordag, U. Mohidden, and V. M. Mostepanenko,
Phys. Rep. {\bf 353}, 1 (2001).
\bibitem{Kirs01} K. Kirsten, {\it Spectral functions in Mathematics
and Physics}. CRC Press, Boca Raton, 2001.
\bibitem{Schw}K. A. Milton, L. L. DeRaad, and
 J. Schwinger, Ann. Phys. (N.Y)115, 338(1978).
\bibitem{setare}M. R. Setare and A. A. Saharain, Int. J. Mod. Phys. {\bf A16},
1463, (2001); M. R. Setare,  Class. Quant. Grav. {\bf 18}, 2097,
(2001); M. R. Setare and R. Mansouri, Class. Quant. Grav. {\bf
18}, 2659, (2001); M. R. Setare,  Class. Quant. Grav. {\bf 18},
4823, (2001);  A. A. Saharian and M. R. Setare, Phys. Lett. {\bf
B552}, 119, (2003);  A. A. Saharian and  M. R. Setare, Phys. Lett.
{\bf B584}, 306, (2004); A. A. Saharian and M. R. Setare, accepted
for publication in Nuclear Physics B.
\bibitem{set}M. R. Setare and R. Mansouri, Class. Quant. Grav. {\bf 18}, 2331,
(2001).
\bibitem{mil}C. M. Bender and K. A. Milton, Phys. Rev. {\bf D55},
6547, (1994).
\bibitem{HHS} J.B.Hartle and S.W.Hawking, Phys. Rev. D. {\bf 28} (1983), 2960. ;
A. D. Sakharov, Sov. Phys.-JETP. {\bf 60} (1984), 214.
\bibitem{CSC}G. F. R. Ellis, A. Sumruk, D. Coule and C. Hellaby, Class.
Quantum Grav. {\bf 9} (1992), 1535. ; S. A. Hayward, Class.
Quantum Grav. {\bf 9} (1992), 1851. ; Class. Quantum Grav. {\bf
10} (1993), L7. ; Phys. Rev. D. {\bf 52} (1995), 7331. ; T.
Dereli, and R. W. Tucker, Class. Quantum Grav. {\bf 10} (1993),
365. ; M. Kossowski and M. Kriele Proc. R. Soc. Lond. A. {\bf 446}
(1995), 115. ; Class. Quantum Grav. {\bf 10} (1993-a), 1157. ;
Class. Quantum Grav. {\bf 10} (1993-b), 2363. ; C. Hellaby and T.
Dray, Phys. Rev. D. {\bf 49} (1994), 5096. ;  J. Math. Phys. {\bf
35} (1994), 5922. ; Phys. Rev. D. {\bf 52} (1995), 7333.
\bibitem{D} T. Dray, C. A. Manogue and R. W. Tucker, Phys. Rev. D. {\bf 48} (1993), 2587. ;
 Gen. Rel. Grav. {\bf 23} (1991), 967. ;  Class. Quantum  Grav. {\bf 12} (1995), 2767. ; J.
D. Romano, Phys. Rev. D, {\bf 47} (1993), 4328. ; J. Gratus and R.
W. Tucker, J. Math. Phys. {\bf 36} (1995), 3353. ; J. Math. Phys.
{\bf 37} (1996), 6018.
\bibitem{FM}F. Darabi, M. R. Setare, {\it Casimir effect in a two dimensional signature changing
 spacetime}, gr-qc/0507043.
\end{thebibliography}
\end{document}